\documentclass[epj]{webofc}
\usepackage[varg]{txfonts}
\usepackage{subfigure}
\usepackage{epsfig}
\usepackage{graphicx}
\usepackage{amssymb}
\usepackage{lineno}
\usepackage{hyperref}   
\wocname{EPJ Web of Conferences}
\woctitle{ICNFP 2017}
\newcommand{\pt}{\ensuremath{p_{\rm T}}}
\begin{document}
\selectlanguage{english}
\title{Measurement of J/$\psi$ production as a function of  multiplicity in pp and p-Pb collisions with ALICE}
%
%

\author{Anisa Khatun\inst{1,2}\fnsep\thanks{\email{anisa.khatun@cern.ch}} for ALICE Collaboration 
}

\institute{Aligarh Muslim University, Aligarh, India
\and
           Saha Institute of Nuclear Physics, Kolkata, India
}

\abstract{%
  The increase of hard probe production as a function of the charged particle multiplicity in proton-proton  and proton-lead collisions is considered to be an interesting observable for the study of multiple parton interactions. In the present work, the correlation between  J/$\psi$ production and  charged particle multiplicity has been reviewed in pp collisions at $\sqrt{s} = 7$ and 13 TeV  and  p-Pb collisions at $\sqrt{s_{\rm NN}}$ = 5.02 TeV  at mid- and forward rapidities. The J/$\psi$ measurement in pp collisions at $\sqrt{s}$ = 13 TeV  using events triggered by the ALICE electromagnetic calorimeter at midrapidity is discussed in this report, too. An increment of the relative J/$\psi$ yields has been observed as a function of the multiplicity. The results have  also been compared to theoretical model predictions.  %
}
\maketitle
%

\section{Introduction}
\label{intro}
The production of charmonium (bound states of c and $\bar{\rm c}$ quarks) in heavy-ion collisions is an ideal probe to explore the Quark Gluon Plasma (QGP) in the laboratory \cite{MSATZ}. The study of quarkonium production in ultra-relativistic proton-proton (pp) collisions provides a  crucial tool for testing  hadronisation models and QCD as well as a baseline for lead-lead (Pb-Pb) measurements. In p-Pb collisions, J/$\psi$ measurements help to investigate Cold Nuclear Matter (CNM) Effects such as gluon shadowing, gluon saturation, coherent parton energy loss or nuclear absorption and also provide a  reference for Pb-Pb measurements \cite{JMB}. 

The multiplicity measurement is important to estimate the general properties of the collision. The hadron production as a function of multiplicity has drawn much attention in the recent time since this study is useful for understanding Multiple Parton Interaction (MPI) and the dependence of particle production on the event multiplicity \cite{AliceA}. MPIs, in general are used to describe the soft processes leading to bulk production, can contribute to the rarer hard processes like quarkonia production. This study can play an important role to understand the production mechanism of soft scale processes and its relation with heavy quarks produced in the hard processes \cite{SARAPort}.

An increase of the production as a function of multiplicity has been observed by ALICE for both the J/$\psi$ and the D mesons \cite{Alice3}. The effects related to MPIs observed in pp might also be relevant in p-Pb collisions, where the number of MPI is proportional to number of binary nucleon-nucleon collisions \cite{AMorsch}.

\section{Experimental setup and analysis strategy}
\label{sec-1}
In the present analysis data from the following sub-systems are used: the central barrel detectors and forward detectors such as Muon Spectrometer (MS), T0 and V0 \cite{AliceC}. The central barrel system consists of Inner Tracking System (ITS), Time Projection Chamber (TPC), Transition Radiation Detector (TRD) and  ElectroMagenetic CALorimeter (EMCAL) sub-detectors, which are placed inside a 0.5 T magnetic field. The ITS consists of six layers of silicon detectors and is located close to the beam axis. The two innermost layers of the ITS are the Silicon Pixel Detectors (SPD) with radii 3.9 $\rm cm$ and  7.6 $\rm cm$ which have been used for multiplicity estimation in this analysis. The charged particle multiplicity has been measured as tracklets produced by hits on the SPD readout chips \cite{Alice2}.

The J/$\psi$ production measurement has been carried out in two rapidity regions, at midrapidity via electron decay mode  (|$\eta$|< 0.9) and at forward rapidity ($-4.0 < \eta < -2.5$) via muon decay mode using the MS. The ITS also provides collision vertex and additional track information for J/$\psi$ reconstruction with $e^{-}e^{+}$ pair at midrapidity ($|y_{\rm lab}|<0.9$). The TPC performs track reconstructions and particle identification via the measurement of specific energy loss. In this report, events with high-\pt\ electrons are selected by a level-1 trigger in the EMCAL, requiring a minimum energy deposit of 7 $\rm GeV$ \cite{emcalAna}. The V0 placed on either side of the interaction point (IP), respectively at $-90 \rm\ cm$ and  $+340 \rm\ cm$ from $ \it{z}$ vertex, is used as a minimum bias trigger detector. The T0 detector, made of two arrays of Cherenkov counters covering pseudorapidity ranges  2.9 < $\eta$ < 3.3  and  $-5 < \eta < -4.5$, is also used for timing an triggering purposes. The MS situated along the negative $z$ axis (in the ALICE reference frame) at a distance of 90  to 1720 $\rm cm$ from the IP, provides acceptance for quarkonia down to zero transverse momentum (\pt\ $\sim 0 $).

The muon tracks have been recorded from the hits at the cathode pad chambers of muon tracking systems with $\sim$10\ $\rm m$ tracking length starting from $\sim$500\ $\rm cm$ from the IP. The triggers for the unlike sign muon pairs have been issued by the Muon Trigger system consisting of four planes of Resistive Plate Chambers arranged in two stations. A muon filter and a front absorber help removing hadrons and secondary particles \cite{RefB}.

\begin{figure}[h!]
\centering
\sidecaption
\includegraphics[width=7.2cm,height=7cm]{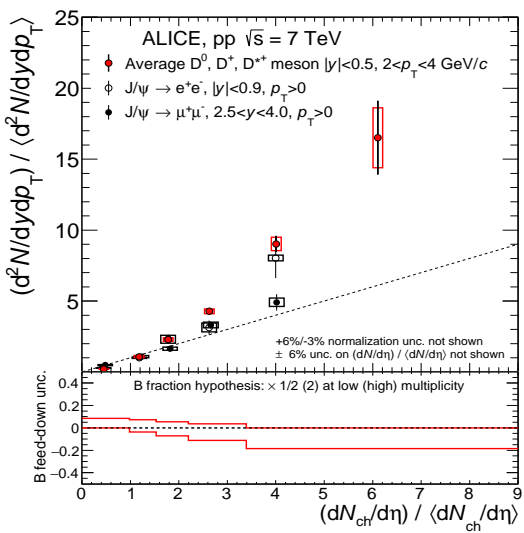}

\caption{ Relative inclusive J/$\psi$ and D meson yield as function of relative charged particle multiplicity in pp collisions at $\sqrt{s} = 7$ TeV at mid- and forward rapidities.}
\label{fig-1}
\end{figure}

\section{J/$\psi$ measurement as a function of multiplicity}
\label{sec-2}
Various studies are performed by ALICE to understand the correlation between soft processes, responsible for bulk multiplicity, and hard processes such as quarkonium production in hadron-hadron and hadron-ion collisions. The dependence of the J/$\psi$ yield as a function of multiplicity was evaluated in pp collisions at $\sqrt{s} = 7$ TeV, along with D mesons, in both mid-rapidity and forward rapidity regions \cite{Alice3, Alice2}. The study showed a linear increment of the relative yield of the J/$\psi$ and the D mesons with increasing charged particle multiplicity as shown in Fig.~\ref{fig-1}. The fact that both D and J/$\psi$ production increase as a function of multiplicity suggests that this is related to charm production rather than to a specific hadronisation process. 

\begin{figure}[h!]
\centering
\includegraphics[width=8cm,height=7cm]{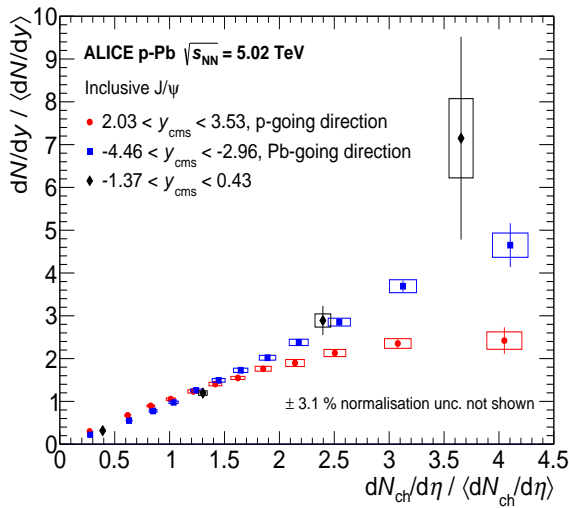}
\caption{ Relative  inclusive J/$\psi$ yield as a function of relative charged particle  multiplicity in 3 rapidity regions in p-Pb collisions at $\sqrt{s_{\rm NN}}$ = 5.02 TeV. }
\label{fig-pb}
\end{figure}

The multiplicity dependence has also been studied in p-Pb collisions at $\sqrt{s_{\rm NN}}$ = 5.02 TeV in forward ($2.03< y_{\rm cms} <3.53$), backward ($-4.46< y_{\rm cms} <-2.96$) and midrapidity ($-1.37< y_{\rm cms} <0.43$) regions, as shown in Fig.~\ref{fig-pb} \cite{Alice4}. An increment of relative J/$\psi$ yield is observed in all three rapidity regions at low multiplicities. The forward rapidity region shows saturation towards higher multiplicities. 

\begin{figure}[h!]
\centering
\subfigure[]{\includegraphics[width=7cm,height=5.8cm]{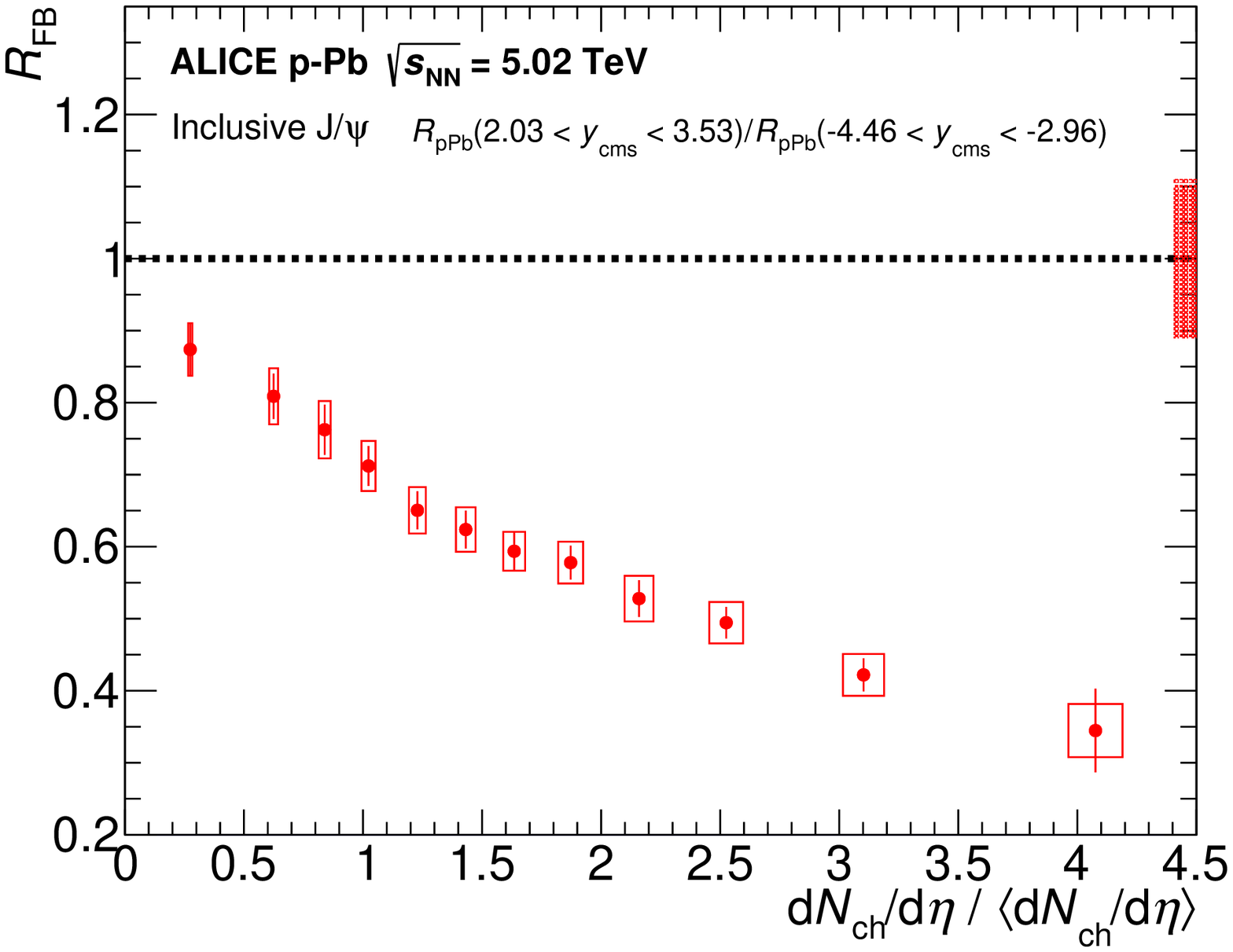}}
\subfigure[]{\includegraphics[width=7cm,height=5.8cm]{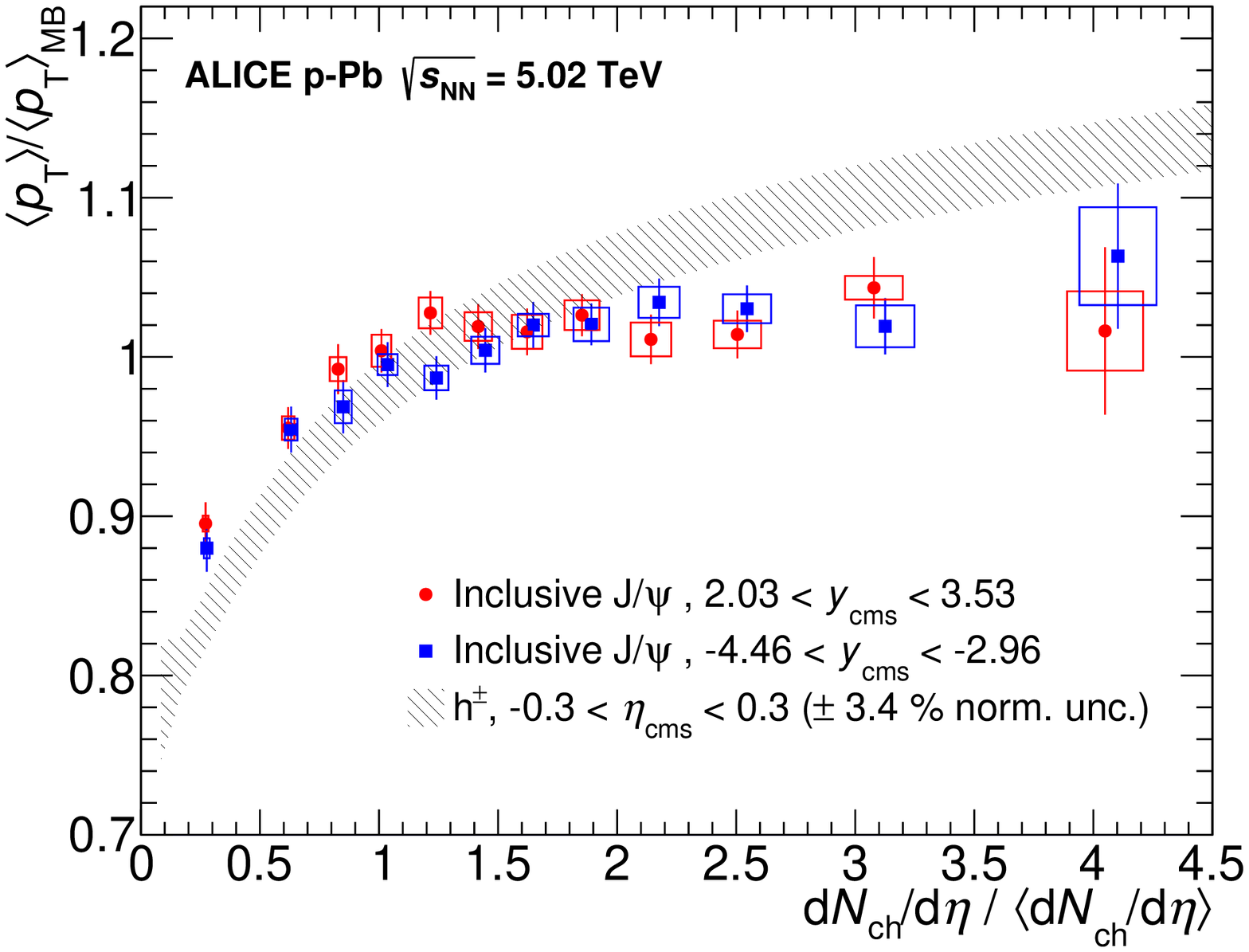}}
\caption{ (a) The ratio $R_{\rm FB}$ of inclusive J/$\psi$ in p-Pb collisions at $\sqrt{s_{\rm NN}}$ = 5.02 TeV as a function of multiplicity. (b) Relative \pt\ for J/$\psi$ as function of relative charged particle multiplicity for forward and backward rapidity.  }
\label{fig-2}       
\end{figure}

The ratio of the yields at forward and backward rapidities $R_{\rm FB}$ as a function of multiplicity is shown in Fig.~\ref{fig-2}. The figure shows a clear suppression of the J/$\psi$ yield at forward rapidity with respect to backward rapidity and the suppression increases with increasing charged particle multiplicity, consistently with predictions including gluon shadowing. The relative average transverse momentum (<\pt>) of J/$\psi$  as a function of charged particle density is reported for forward and backward rapidities in Fig.~\ref{fig-2}, the same trend is observed for the two rapidity ranges \cite{Alice4}. The dashed lines represents the $\langle \pt \rangle$ of charged hadrons as a function of multiplicity in the midrapidity region.

\begin{figure}[h!]
\centering
\subfigure[]{\includegraphics[width=6.8cm,height=5.6cm]{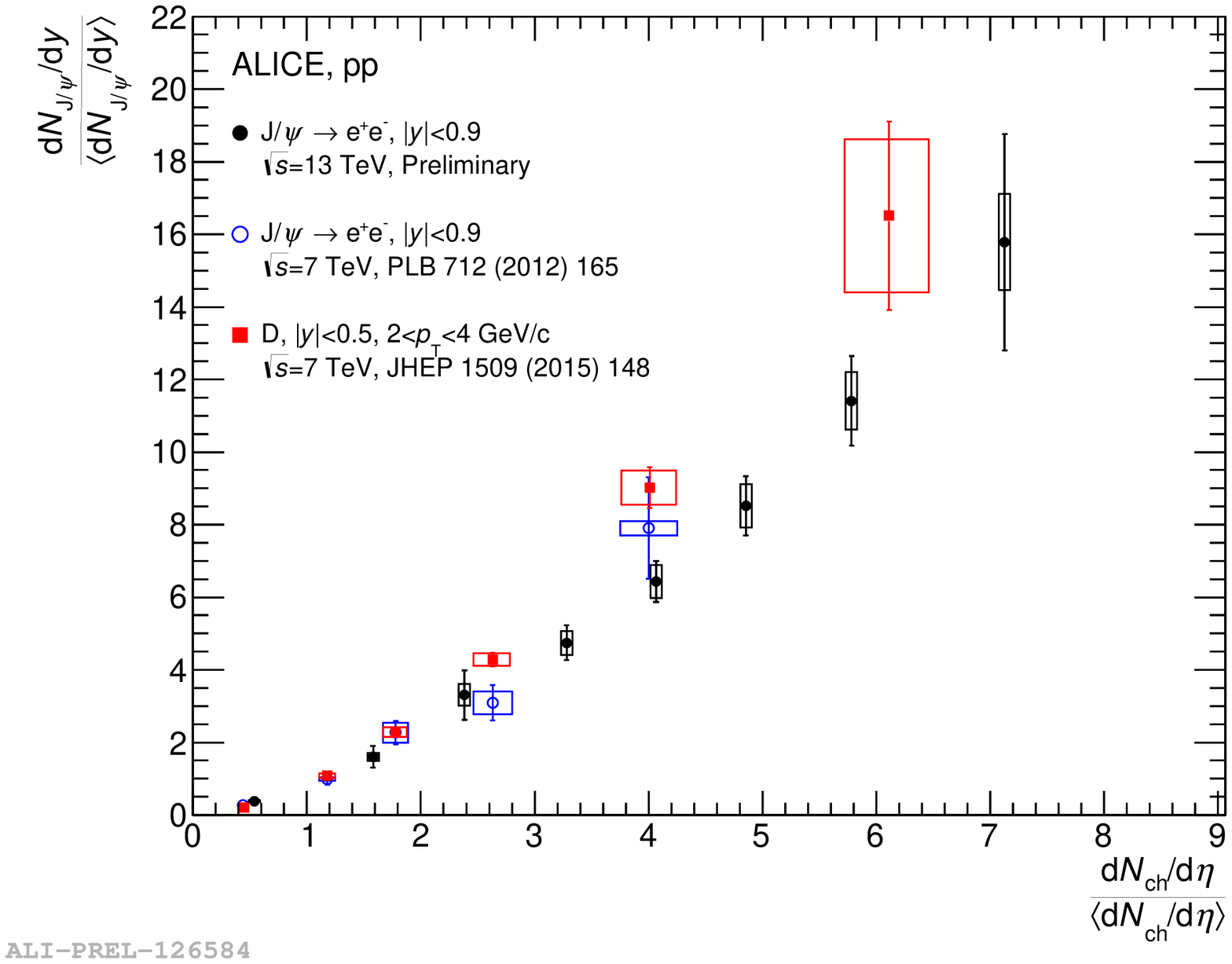}}
\subfigure[]{\includegraphics[width=6.8cm,height=5.6cm]{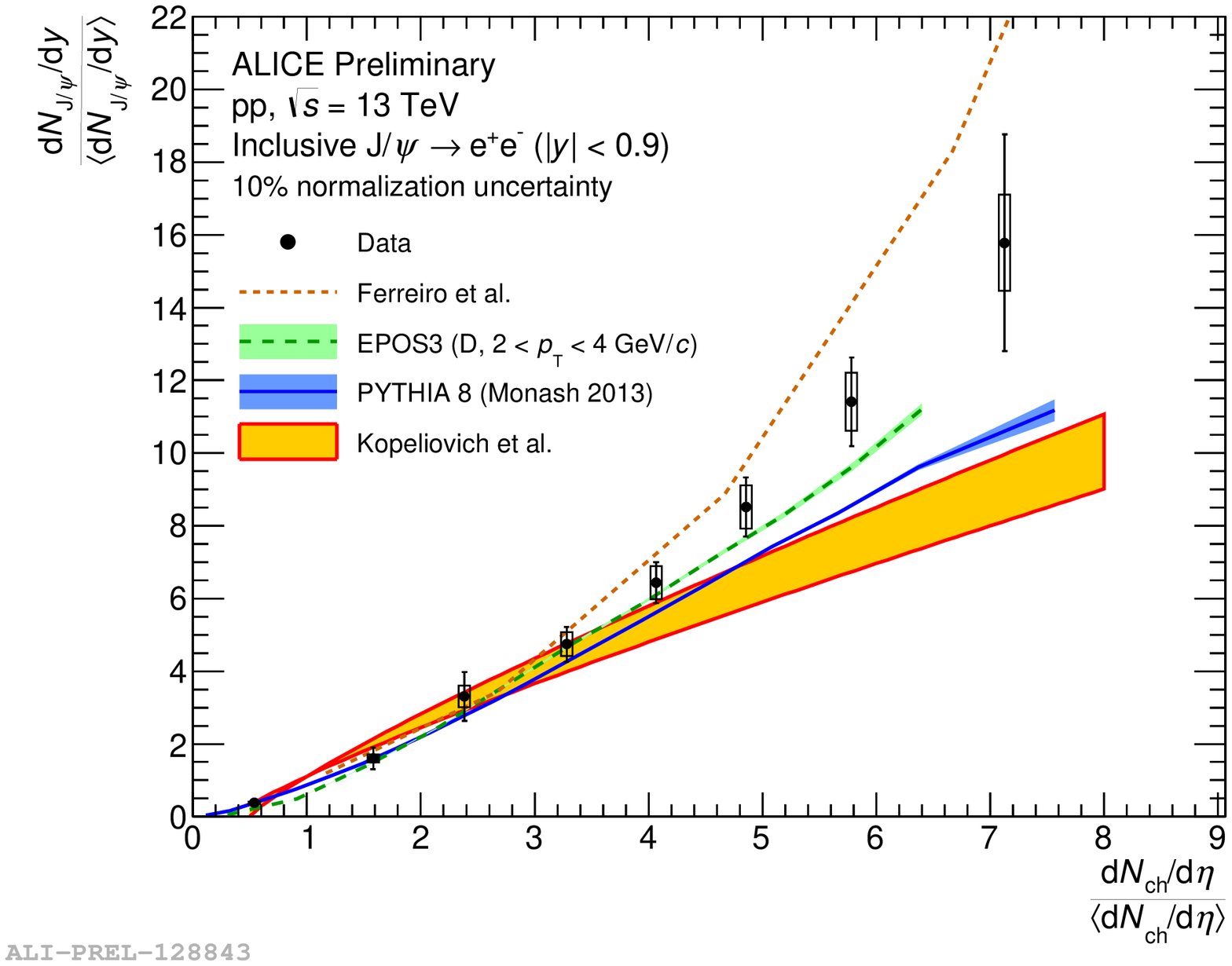}}
\caption{(a) Comparison of relative J/$\psi$ yield  at $\sqrt{s} = 7$ TeV, 13 TeV and D mesons yield at $\sqrt{s} = 7$ TeV as function of relative charged particle multiplicity in pp collisions. (b) Dependence of relative J/$\psi$ as a function of multiplicity in pp collisions at $\sqrt{s} = 13$ TeV compared with theoretical models. }
\label{fig-3}       
\end{figure}

The same study has been carried out in pp collisions at $\sqrt{s} = 13$	TeV in the midrapidity region and compared with the previous results obtained by ALICE in pp collisions, and is shown in Fig.~\ref{fig-3} \cite{Alice13tev}. The multiplicity range is much larger, spanning twice the range explored in the previous study, showing similar trend with a slope significantly larger than unity. These results have been compared to predictions from the available theoretical models  (Fig.~\ref{fig-3}) by Ferreiro \cite{Alicemodels1}, EPOS3 \cite{Alicemodels2}, PYTHIA8 \cite{Alicemodels3}, Kopeliovich \cite{Alicemodels4} in pp collisions at $\sqrt{s} = 13$ TeV. All models predict a J/$\psi$ enhancement with multiplicity. However, at higher multiplicity values the models show different trends. EPOS3 model is in fair agreement with the data. 

\begin{figure}[h!]
\centering
\subfigure[]{\includegraphics[width=7cm,height=6cm]{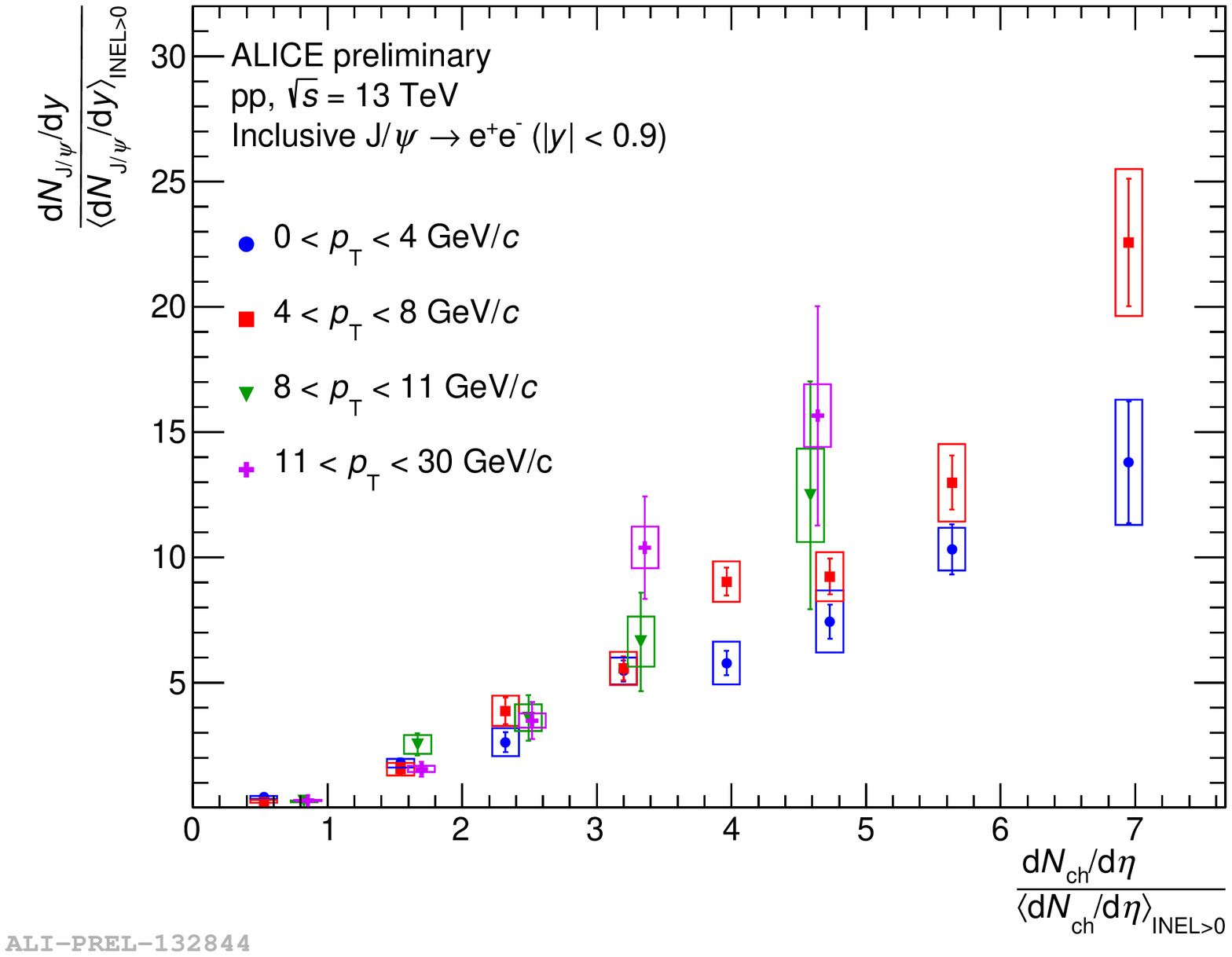}}
\subfigure[]{\includegraphics[width=7cm,height=6cm]{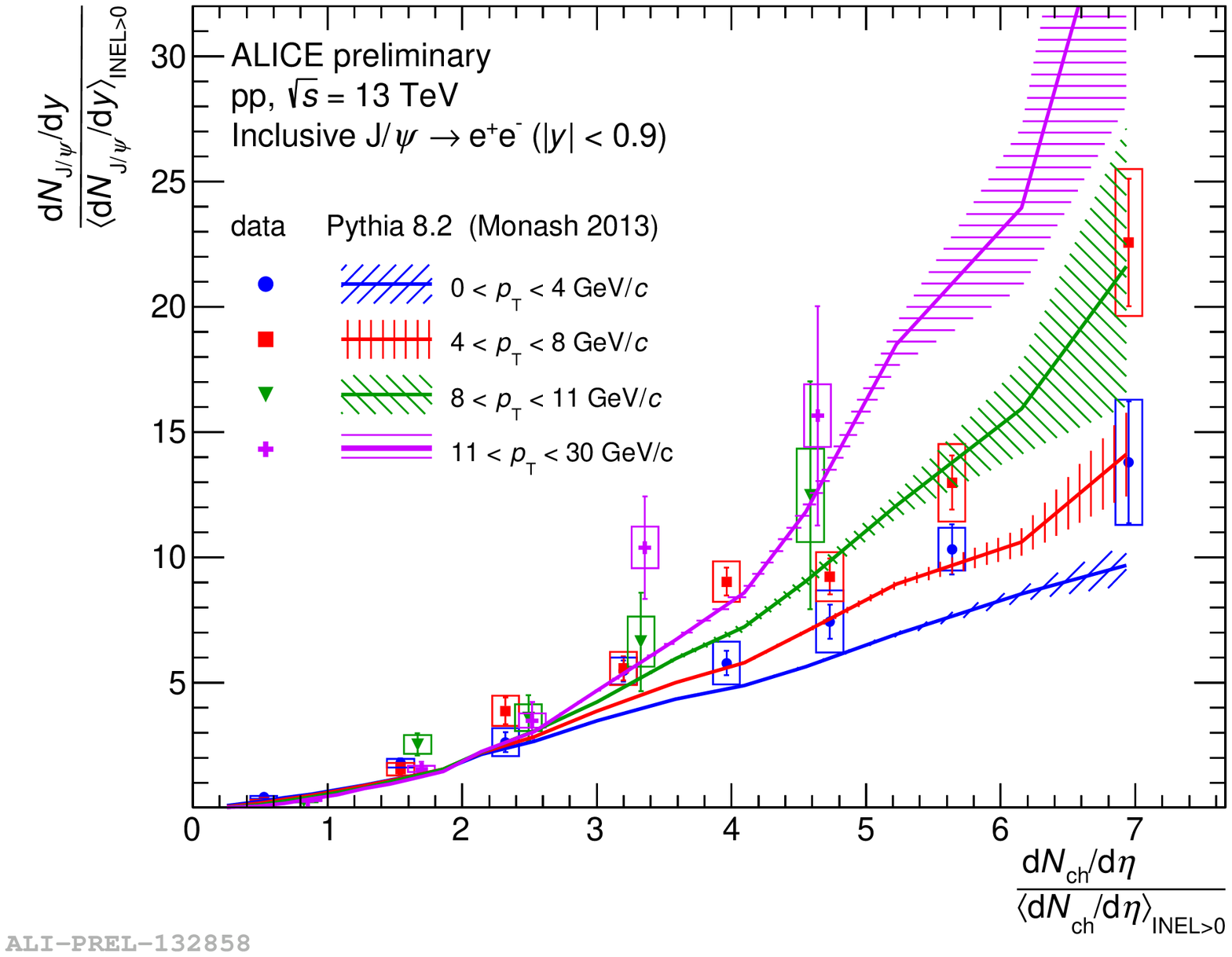}}
\caption{(a) Normalised J/$\psi$ yield as a function of relative charged particle multiplicity in pp collisions at $\sqrt{s} = 13$ TeV with EMCAL in four different \pt\ bins. (b) The relative J/$\psi$ yield as a function of multiplicity in pp collisions at $\sqrt{s} = 13$ TeV compared to PYTHIA8.2 (Monash 2013) in four different \pt\ bins. }
\label{fig-emcal}       
\end{figure}

The EMCAL triggered events allowed us to study high \pt\ as well as high multiplicity events. ALICE measured the J/$\psi$ normalized yield as a function of the normalized charged particle multiplicity up to 30 $\mathrm{GeV}$ \pt\  divided in four \pt\ bins as shown in Fig. \ref{fig-emcal}. This analysis indicates stronger than linear increase of J/$\psi$ yield as function of multiplicity with increasing transverse momentum. Results obtained using PYTHIA8.2 as event generator are in agreement with experimental data in the defined \pt\ intervals.

\section{Conclusions}
The multiplicity dependence of J/$\psi$  production in pp and p-Pb collisions at different energies and in different rapidity regions has been presented. The D meson yield is also reported at midrapidity in pp collisions at $\sqrt{s}$ = 7 TeV. The increases of J/$\psi$  with multiplicity are similar for all the considered collision energies and rapidity ranges. D mesons also show the same trend. \\
The measurement of J/$\psi$ production relative to multiplicity in p-Pb collisions at $\sqrt{s_{\rm NN}}$ = 5.02 TeV is explored in three rapidity regions, showing saturation at the forward rapidity. The relative <\pt>  J/$\psi$ yield shows saturation above the relative multiplicity range > 1.5. \\
The relative J/$\psi$ yield increases steeply with multiplicity in pp at 13 TeV and the effect is stronger for high \pt. The larger statistics collected in Run-2 at LHC allowed us to extend the measurement to larger charged particle multiplicities and to higher \pt. The experimental results are qualitatively described by the existing theoretical models. It will be interesting to explore the energy dependence of the J/$\psi$ production as a function of charged particle multiplicity over a wider centre of mass energy range. In ALICE such studies will be performed at the currently available collision energies namely, in pp collisions at $\sqrt{s}$ = 2.76, 5.02  and 13 TeV and in p-Pb collsions at $\sqrt{s_{\rm NN}}$ = 8.16 TeV.

%
%
%


\end{document}